\documentclass[%
 12pt,
 reprint,
  twocolumn,
 showpacs,
 showkeys,
 preprintnumbers,
 amsmath,amssymb,
 aps,
  pra,
  longbibliography,
 ]{revtex4-2}

\usepackage[breaklinks=true,colorlinks=true,anchorcolor=blue,citecolor=blue,filecolor=blue,menucolor=blue,pagecolor=blue,urlcolor=blue,linkcolor=blue]{hyperref}
\usepackage{url}

\usepackage{graphicx}

\usepackage[usenames,dvipsnames]{xcolor}

\usepackage{musixtex}

\begin{document}


\title{Quantum music, quantum arts and their perception}

\author{Volkmar Putz}
\affiliation{P\"adagogische Hochschule Wien, Grenzackerstra\ss e 18, A-1100
    Vienna, Austria}
\email{volkmar.putz@phwien.ac.at}

\author{Karl Svozil}
\affiliation{Institute for Theoretical Physics, Vienna
    University of Technology, Wiedner Hauptstra\ss e 8-10/136, A-1040
    Vienna, Austria}
\email{svozil@tuwien.ac.at} \homepage[]{http://tph.tuwien.ac.at/~svozil}

\pacs{03.65.Aa, 03.67.Ac}
\keywords{music, quantum theory, field theory, piano}

\begin{abstract}
The expression of human art, and supposedly sentient art in general, is modulated by the available rendition, receiving and communication techniques. The components or instruments of these techniques ultimately exhibit a physical, in particular, quantum layer, which in turn translates into physical and technological capacities to comprehend and utilize what is possible in our universe. In this sense, we can apply a sort of Church-Turing thesis to art, or at least to its rendition.
\end{abstract}

\maketitle

\section{Realm of quantum expressibility}
\label{2021-qart-s1}

A short glance at historic practices of music and artistic expression, in general, suggests that there has been, and still is,
a fruitful exchange of ideas between craftsmanship, technology, and (material) sciences on the one hand,
and entertainment, artistry, and creativity on the other hand.
Impulses and ideas flow back and forth, very much like in the accompanying fields of mathematics and natural sciences.
This is even true, in particular, for culinary subjects such as molecular gastronomy,
where it has been argued that ``food processing dominates cuisines'': because even if all of the French recipes
would have been erased from people and other memories, most if not all of these revered dishes
could be ``recovered'' by merely following ``reasonable'' rules of food processing~\cite{this-herve-bjn}---which
strongly are linked to technology, such as the ``domestication'' of fire.

It thus comes of no surprise that the evolution of quantum physics brought about
the quest for the quantum arts; and in particular, for quantum music~\cite{2015-qmusic}
and quantum fine arts, especially quantum visual art.
Indeed, every aspect of human life can be re-evaluated and reframed in terms of the quantum paradigm.

In our (not so humble) opinion there are two immediate issues:
One issue is the tendency to re-introduce irrational ``magic'', a sort of
``quantum hocus pocus''~\cite{svozil-2016-quantum-hokus-pokus}
that brings it close to the esoteric, and fosters a kind of pseudo-religion allegedly
justified by the most advanced contemporary physics.

Another, converse, issue is the temptation to argue that, just like in quantum computing~\cite[Section~1.1]{mermin-07},
``any art is quantum''
as the ``underlying physical layer'' of any (classical) artistic expression is governed by the laws of quantum mechanics.
However, we emphasize upfront that we have to resist this temptation towards a premature
flattening and folding of the quantum phenomena into classical molds.
Rather we consider quantum arts, and, in particular, quantum music,
as  operations exploiting certain very special transformations
of physical internal states, subject to very carefully controlled conditions.

So what exactly are these very special transformations that characterize quantum art?
In this regard, we can proceed almost in parallel to the development of quantum computation~\cite{fortnov-03,nielsen-book10,mermin-07},
and point out some central assets or capacities:
\begin{itemize}
\item[(i)] parallelization through coherent superposition (aka simultaneous linear combination) of classically mutually exclusive
tones or signals that are acoustic,  optic, touch, taste,  or otherwise sensory;
\item[(ii)] entanglement not merely by classical correlation~\cite{peres222}
but by relational encoding~\cite{schrodinger,zeil-bruk-99a,zeil-99,zeil-Zuk-bruk-01}
of multi-partite states such that any classical information is ``scrambled'' into relational, joint multi-partite properties
while at the same time losing value definiteness about the single constituents of such multi-partite states ---this
can be seen as a sort of zero-sum game, a tradeoff between individual and collective properties;
\item[(iii)]
complementarity associated with value (in)definiteness of certain tones or signals
that is acoustic,  optic, touch, taste,  or otherwise:
if one such observable is definite, another is not, and \textit{vice versa};
\item[(iv)]
contextuality is an ``enhanced'' form of complementarity and value indefiniteness that can be defined in various
ways~\cite{Dzhafarov-2017,Abramsky2018,Grangier_2002,Auffeves-Grangier-2018,Auffves2020,Grangier-2020,cabello2021contextuality},
in particular, emphasizing homomorphic, structure-preserving nonembeddability into classical schemes~\cite{specker-60,kochen1,svozil-2021-context}
\end{itemize}

Those criteria or observables constitute significant signatures of quantum behavior.
The transformations and processing of classical-to-quantum states or quantum states exhibiting these features
can be considered musical, optical, or other instruments or ``transmitters'' for the creation of quantum art.
Similarly, assorted transformations process quantum art.
Finally, the process of information transmission requires instruments of perception or ``receivers''~\cite[Fig.~1]{shannon}.

Let us mention typical components and theoretical entities as example transformations.
For instance, Hadamard transformations produce perfect ``mixtures'' of classically mutually exclusive signals.
Quantum Fourier transforms produce generalized mixtures.
All of them have to be uniformly unitary---that is,
in terms of the various equivalent formal definitions, they have to transform orthonormal basis into orthonormal ones,
they have to preserve scalar products or norms, and their inverse is the adjoint.
One of the physical realizations is in terms of generalized beam splitters~\cite{rzbb,zukowski-97}.

Depending on whether we are willing to contemplate genuine quantum receivers or merely classical ones we
end up with either a quantum cognition or with merely a classical cognition of this quantum art; and,
in particular, of quantum music.
In the first, radical deviation from classical music, we would have to accept the possibility of
human or sentient consciousness or audience to perceive quantum impressions.

This is ultimately a neurophysiologic question.
It might well be that the processing of signals exterior to the receiving and perceiving
``somewhere along those channels'' requires a breakdown to classicality; most likely
through the introduction of stochasticity~\cite{Glauber-cat-86}.
This is very much in the spirit of Schr\"odinger's cat~\cite{schrodinger}
and (later) quantum jellyfish~\cite{schroedinger-interpretation} metaphors
based on the assumption that, ultimately, even if decoherence by environmental intake
can be controlled, there cannot be any simultaneous co-experience of being both dead and alive,
just as there might not be any co-experience of passing into a room by two separate doors simultaneously.

On the other hand, nesting of the Wigner's friend  type~\cite{v-neumann-49,everett,wigner:mb,everett-collw}, suggests that
there might be substance to a sort of mindful co-experience of two classical distinct experiences.
Whether such experiences remain on the subconscious primordial level of perception,
or whether this can be levied to a full cognitive level is a fascinating question on its own that exceeds the limited scope of this article.

\section{Quantum musical tones}
\label{2021-qart-s2}

In what follows we closely follow our nomenclature and presentation of quantum music~\cite{2015-qmusic}.
Those formal choices are neither unique nor comprehensive.
Alternatives are mentioned.

We consider a quantum octave in the C major scale, which classically consists
of the tones  $c$, $d$, $e$, $f$, $g$, $a$, and $b$,
represented by eight consecutive white keys on a piano. (Other scales are straightforward.)
At least three ways to quantize this situation can be given:
\begin{itemize}
\item[(i)] bundling octaves by coherent their superposition (aka simultaneous linear combination), as well as
\item[(ii)] considering pseudo-field theoretic models
treating notes as field modes that are either bosonic or fermionic.
\end{itemize}

The seven tones
$c$, $d$, $e$, $f$, $g$, $a$, and $b$ of the octave can be considered as belonging to disjoint events
(maybe together with the null event $0$) whose probabilities should add up to unity.
This essentially suggests a formalization by a seven (or eight) dimensional Hilbert space $\mathbb{C}^7$  or $\mathbb{C}^8$)
with the standard Euclidean scalar product.
The respective Hilbert space represents a full octave.

\begin{figure}
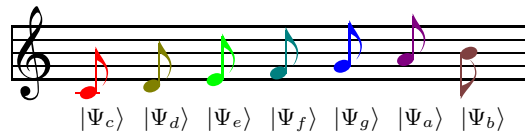

\begin{center}
\begin{music}
\startextract
\NOtes\zsong{$\vert \Psi_c \rangle$}{\color[rgb]{1,0,0}\ca{c}}\en
\NOtes\zsong{$\vert \Psi_d \rangle$}{\color[rgb]{0.5,0.5,0}\ca{d}}\en
\NOtes\zsong{$\vert \Psi_e \rangle$}{\color[rgb]{0,1,0}\ca{e}}\en
\NOtes\zsong{$\vert \Psi_f \rangle$}{\color[rgb]{0,0.5,0.5}\ca{f}}\en
\NOtes\zsong{$\vert \Psi_g \rangle$}{\color[rgb]{0,0,1}\ca{g}}\en
\NOtes\zsong{$\vert \Psi_a \rangle$}{\color[rgb]{0.5,0,0.5}\ca{h}}\en
\NOtes\zsong{$\vert \Psi_b \rangle$}{\color[rgb]{0.5,0.25,0.25}\ca{i}}\en
\zendextract
\end{music}
\end{center}
\caption{(Color online) Temporal succession of quantum tones
$\vert \Psi_c \rangle$,
$\vert \Psi_d \rangle$,~$\ldots$,
$\vert \Psi_b \rangle$
in the C major scale
forming the octave basis ${\mathfrak B}$.}
\label{2015-qmusic-fig1}
\end{figure}

We shall study the seven-dimensional case $\mathbb{C}^7$.
The seven tones forming one octave can then be represented as an orthonormal basis ${\mathfrak B}$
of  $\mathbb{C}^7$ by forming the set theoretical union of the mutually orthogonal
unit vectors; that is, ${\mathfrak B} = \{ \vert \Psi_c \rangle , \vert \Psi_d \rangle,\ldots \vert \Psi_b \rangle \}$,
where the basis elements are the Cartesian basis tuples
\begin{align*}
\vert \Psi_c \rangle &=\begin{pmatrix}0,0,0,0,0,0,1\end{pmatrix}, \\
\vert \Psi_d \rangle &=\begin{pmatrix}0,0,0,0,0,1,0\end{pmatrix},\\
&\ldots                                                           \\
\vert \Psi_b \rangle &=\begin{pmatrix}1,0,0,0,0,0,0\end{pmatrix}
\end{align*}
of $\mathbb{C}^7$.
Fig.~\ref{2015-qmusic-fig1} depicts the basis ${\mathfrak B}$ by its elements, drawn in different colors.

\subsection{Bundling octaves into single tones}

Pure quantum musical states could be represented as unit vectors
$\vert \psi \rangle \in \mathbb{C}^7$
which are linear combinations of the basis ${\mathfrak B}$; that is,
\begin{equation}
\vert \psi \rangle =
\alpha_c \vert \Psi_c \rangle
+ \alpha_d \vert \Psi_d \rangle
+\cdots
+\alpha_b \vert \Psi_b \rangle
,
\label{2015-qmusic-e1}
\end{equation}
with coefficients $\alpha_i$
satisfying
\begin{equation}
\vert \alpha_c  \vert^2
+
\vert \alpha_d  \vert^2
+ \cdots  +
\vert \alpha_b  \vert^2
=1
.
\end{equation}
Equivalent representations of $\vert \psi \rangle$ are in terms of
the
one-dimensional subspace
$\{ \vert \phi \rangle  \mid \vert \phi \rangle = \alpha \vert \psi \rangle ,\, \alpha \in \mathbb{C}\}$
 spanned by $\vert \psi \rangle$, or by the projector
$\textsf{\textbf{E}}_\psi =  \vert \psi \rangle  \langle \psi  \vert$.

A musical ``composition''---indeed, and any
 succession of quantized tones forming
a ``melody''---would be obtained by successive unitary permutations of the state $\vert \psi \rangle$.
The realm of such compositions would be spanned by the succession of all
unitary transformations $\textsf{\textbf{U}}: {\mathfrak B} \mapsto {\mathfrak B}'$
mapping some orthonormal basis ${\mathfrak B}$ into another orthonormal basis ${\mathfrak B}'$;
that is~\cite{Schwinger.60}, $\textsf{\textbf{U}} =   \sum_i \vert {\Psi'}_i \rangle  \langle \Psi_i  \vert$.

\subsection{Coherent superposition of tones as a new form of musical parallelism}

\begin{figure}
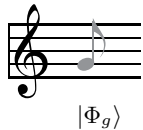

\begin{center}
\begin{music}
\startextract
\NOtes\zsong{$\vert \Phi_g \rangle$}{\color{Black!50}\ca{g}}\en
\zendextract
\end{music}
\end{center}
\caption{Representation of a 50:50 quantum tone $\vert \Phi_g \rangle =
\frac{1}{\sqrt{2}}
\left(
\vert 0_g \rangle
-
\vert 1_g \rangle
\right)
$ in gray (without indicating phase factors).}
\label{2015-qmusic-fig3}
\end{figure}

One of the mind-boggling quantum features of this ``bundling'' is the possibility of the simultaneous ``co-existence''
of classically excluding musical states,
such as a 50:50 quantum $g$  in the C major scale obtained by sending $\vert 0_g \rangle $ through the Hadamard gate
$\textsf{\textbf{H}}= \frac{1}{\sqrt{2}}\begin{pmatrix} 1 & 1 \\ 1 & -1 \end{pmatrix}$,
resulting in
$
\frac{1}{\sqrt{2}}
\left(
\vert 0_g \rangle
-
\vert 1_g \rangle
\right)
$, and depicted in Fig.~\ref{2015-qmusic-fig3} by a 50 white 50 black; that is, gray, tone (though without the relative ``$-$'' phase).

This novel form of musical expression might contribute to novel musical experiences; in particular, if
any such coherent superposition can be perceived by the audience in full quantum uniformity.
This would require the cognition of the recipient to experience quantum coherent superpositions---a capacity
that is highly speculative.
It has been mentioned earlier that any such capacity is related to
Schr\"odinger's cat~\cite{schrodinger} and quantum jellyfish~\cite{schroedinger-interpretation} metaphors, as well as to
nestings of the Wigner's friend type~\cite{v-neumann-49,everett,wigner:mb,everett-collw}.

\subsection{Classical perception of quantum musical parallelism}

In the following, we shall assume that quantum music is ``reduced'' to the continuous infinity of its classical forms.
Then, if a classical auditorium listens to the quantum musical state  $\vert \psi \rangle $ in Eq.~\ref{2015-qmusic-e1},
the individual classical listeners  may perceive $\vert \psi \rangle $ very differently;
that is, they will hear only a {\em single one} of the different tones with probabilities of $
\vert \alpha_c  \vert^2
$, $
\vert \alpha_d  \vert^2
$, $\ldots$, and $
\vert \alpha_b  \vert^2
$, respectively.

Indeed, suppose that classical recipients (aka ``listeners'')
modeled by  classical measurement devices acting as
information-theoretic
receivers are assumed.
Then any perception (aka ``listening'' or reception)  of
a quantum musical state that is in a coherent superposition---with
some coefficients $0<\vert \alpha_i \vert <1$---because of the
supposedly irreducably stochastic~\cite{zeil-05_nature_ofQuantum}
quantum-to-classical translation~\cite{svozil-2003-garda}
represents an ``irreducible''~\cite{PhysRevD.22.879,PhysRevA.25.2208,greenberger2,Nature351,Zajonc-91,PhysRevA.45.7729,PhysRevLett.73.1223,PhysRevLett.75.3783,hkwz}
stochastic measurement.
This can never render a  unique classical listening experience,
as the probability to hear the tone $i$ is $\vert \alpha_i \vert^2$.
Therefore, partitions of the audience will hear different manifestations of the quantum musical
composition made up of all varieties of successions of tones.
These experiences multiply and diverge as more tones are perceived.

\begin{figure}
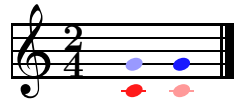

\begin{center}
\begin{music}
\generalmeter{\meterfrac24}
\startextract 
\Notes
{\color[rgb]{1,0.1,0.1}\zq c}{\color[rgb]{0.6,0.6,1}\zq g}  \enotes
\Notes
{\color[rgb]{1,0.6,0.6}\zq c}{\color[rgb]{0.1,0.1,1}\zq g}  \enotes
\Endpiece
\zendextract 
\end{music}
\end{center}
\caption{(Color online) A two-note quantum musical composition---a natural fifth.}
\label{2015-qmusic-fig1a}
\end{figure}

For the sake of a demonstration,
let us try a two-note quantum composition.
We start with a pure quantum mechanical state
in the two-dimensional subspace spanned by  $\vert \Psi_c \rangle $ and $\vert \Psi_g \rangle$,
specified by
\begin{equation}
\vert \psi_1\rangle =
\frac45 \vert \Psi_c \rangle
+ \frac35 \vert \Psi_g \rangle = \frac15 \begin{pmatrix} 4 \\ 3  \end{pmatrix}
.
\end{equation}
$\vert \psi_1 \rangle$ would be detected by the listener as $c$ in 64\%
of all measurements (listenings), and as $g$ in 36\%
of all listenings.
Using the unitary transformation $\textsf{\textbf{X}}= \begin{pmatrix} 0 & 1 \\ 1 & 0 \end{pmatrix}$, the next quantum tone would be
\begin{equation}
\vert \psi_2 \rangle = \textsf{\textbf{X}}  \vert \psi_1 \rangle =
\frac35 \vert \Psi_c \rangle
+ \frac45 \vert \Psi_g \rangle = \frac15 \begin{pmatrix} 3 \\ 4  \end{pmatrix}.
\end{equation}
This means for the quantum melody of both quantum tones $\vert \psi_1\rangle$ and $\vert  \psi_2 \rangle$ in
succession---for the score, see Fig.~\ref{2015-qmusic-fig1a}---that in repeated measurements,
in $0.64^2 = 40.96\%$
of all cases $c-g$ is heard,
in $0.36^2 = 12.96\%$
of all cases $g-c$,
in $0.64\cdot0.36 = 23.04\%$
of all cases $c-c$ or $g-g$, respectively.

\section{Quantum musical entanglement}

Quantum entanglement~\cite{schrodinger} is the property of multipartite quantum systems
to code information ``across quanta'' in such a way that the state of any individual quantum
remains irreducibly indeterminate;
that is, not determined by the entangled multipartite state~\cite{schrodinger,zeil-bruk-99a,zeil-99,zeil-Zuk-bruk-01}.
Thus the entangled whole should not be thought of as composed of its proper parts.
Formally, the composite state cannot be expressed as a product of separate states of the individual quanta.

\begin{figure}
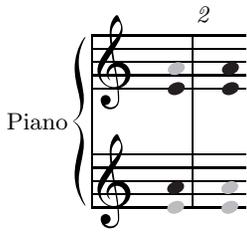

\begin{center}
\begin{music}
\parindent10mm
\instrumentnumber{1} 
\setname1{Piano} 
\setstaffs1{2} 
\startextract 
\notes
\hspace{1 mm}{\color{Black!30}\zq e}{\color{Black!100}\zq h}
|
\hspace{0 mm}{\color{Black!100}\zq e}{\color{Black!30}\zq h}
\en
\bar
\notes
\hspace{1 mm}{\color{Black!30}\zq e}{\color{Black!30}\zq h}
|
\hspace{0 mm}{\color{Black!100}\zq e}{\color{Black!100}\zq h}
\en
\zendextract 
\end{music}
\end{center}
\caption{Quantum musical entangled states
$\vert \Psi_{ea}^- \rangle$ and $\vert \Psi_{ea}^+ \rangle$
in the first bar,
and
$\vert \Phi_{ea}^- \rangle$ and $\vert \Phi_{ea}^+ \rangle$
in the second bar
(without relative phases).}
\label{2015-qmusic-fig4}
\end{figure}

A typical example of an entangled state is the
{\em Bell state},
$\vert \Psi^- \rangle$
or, more generally, states in the Bell basis spanned by the quantized notes $e$ and $a$; that is
\begin{equation}
\begin{split}
\vert \Psi^\pm \rangle = \frac{1}{\sqrt{2}}\left(\vert 0_e \rangle \vert 1_a \rangle \pm \vert 1_e \rangle \vert 0_a \rangle  \right),\\
\vert \Phi^\pm \rangle = \frac{1}{\sqrt{2}}\left(\vert 0_e \rangle \vert 0_a \rangle \pm \vert 1_e \rangle \vert 1_a \rangle  \right),\\
\end{split}
\label{2014-m-ch-fdvs-bellbasis}
\end{equation}
A necessary and sufficient condition
for entanglement among the quantized notes $e$ and $a$ is that the coefficients
$\alpha_1$,
$\alpha_2$,
$\alpha_3$,
$\alpha_4$
of their general composite state
$
\vert \Psi_{ga} \rangle =
\alpha_1 \vert 0_e \rangle \vert 0_a \rangle  +
\alpha_2 \vert 0_e \rangle \vert 1_a \rangle  +
\alpha_3 \vert 1_e \rangle \vert 0_a \rangle  +
\alpha_4 \vert 1_e \rangle \vert 1_a \rangle
$
obey $\alpha_1 \alpha_4 \neq \alpha_2 \alpha_3$~\cite[Sec.~1.5]{mermin-07}.
This is clearly satisfied by Eqs.~(\ref{2014-m-ch-fdvs-bellbasis}).
Fig.~\ref{2015-qmusic-fig4} depicts the entangled musical Bell states.

\begin{figure}
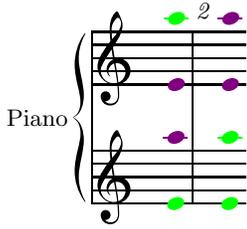

\begin{center}
\begin{music}
\parindent10mm
\instrumentnumber{1} 
\setname1{Piano} 
\setstaffs1{2} 
\startextract 
\notes
\hspace{1 mm}{\color[rgb]{0,1,0}\zq e}{\color[rgb]{0.5,0,0.5}\zq o}
|
\hspace{0 mm}{\color[rgb]{0.5,0,0.5}\zq e}{\color[rgb]{0,1,0}\zq o}
\en
\bar
\notes
\hspace{1 mm}{\color[rgb]{0,1,0}\zq e}{\color[rgb]{0,1,0}\zq o}
|
\hspace{0 mm}{\color[rgb]{0.5,0,0.5}\zq e}{\color[rgb]{0.5,0,0.5}\zq o}
\en
\zendextract 
\end{music}
\end{center}
\caption{(Color online) Quantum musical entangled states for bundled octaves
$\vert \Psi_{ea'}^- \rangle$ and $\vert \Psi_{ea'}^+ \rangle$
in the first bar,
and
$\vert \Phi_{ea'}^- \rangle$ and $\vert \Phi_{ea'}^+ \rangle$
in the second bar
(without relative phases).}
\label{2015-qmusic-fig5}
\end{figure}

Entanglement between different octaves can be constructed similarly.
Fig.~\ref{2015-qmusic-fig5} depicts this configuration for an entanglement between $e$ and $a'$.

\section{Quantum musical complementarity and contextuality}

\begin{figure}
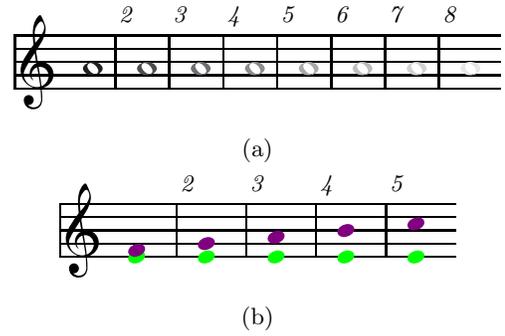

\begin{center}
\begin{music}
\startextract 
\notes
{\color{Black!100}\wh{h}}\enotes\bar
\notes {\color{Black!85}\wh{h}}\enotes\bar
\notes {\color{Black!70}\wh{h}}\enotes\bar
\notes {\color{Black!55}\wh{h}}\enotes\bar
\notes {\color{Black!40}\wh{h}}\enotes\bar
\notes {\color{Black!30}\wh{h}}\enotes\bar
\notes {\color{Black!20}\wh{h}}\enotes\bar
\notes {\color{Black!10}\wh{h}}
\enotes
\zendextract
\end{music}
(a)
\begin{music}
\startextract 
\Notes
{\color[rgb]{0,1,0}\zq e}{\color[rgb]{0.5,0,0.5}\zq f}  \enotes
\bar
\Notes
{\color[rgb]{0,1,0}\zq e}{\color[rgb]{0.5,0,0.5}\zq g}  \enotes
\bar
\Notes
{\color[rgb]{0,1,0}\zq e}{\color[rgb]{0.5,0,0.5}\zq h}  \enotes
\bar
\Notes
{\color[rgb]{0,1,0}\zq e}{\color[rgb]{0.5,0,0.5}\zq i}  \enotes
\bar
\Notes
{\color[rgb]{0,1,0}\zq e}{\color[rgb]{0.5,0,0.5}\zq j}  \enotes
\zendextract 
\end{music}
(b)
\end{center}
\caption{Temporal succession of complementary
tones (a) for binary occupancy $\vert \phi_a \rangle = \alpha_a \vert 0_a \rangle + \beta_a \vert 1_a \rangle$,
with $\vert \alpha_a \vert^2  + \vert \beta_a \vert^2  =1$ with increasing $\vert \alpha_a \vert$ (decreasing occupancy),
(b) in the bundled octave model,
separated by bars.}
\label{2015-qmusic-fig6}
\end{figure}

Although complementarity~\cite{pauli:1933} is mainly discussed in the context of observables,
we can present it in the state formalism by observing that,
as mentioned earlier, any pure state
 $\vert \psi \rangle$ corresponds to the projector
$\textsf{\textbf{E}}_\psi =  \vert \psi \rangle  \langle \psi  \vert$.
In this way,
any two nonvanishing nonorthogonal and noncollinear states
$\vert \psi \rangle$
and
$\vert \phi \rangle$
with $0< \vert \langle \phi \vert \psi \rangle \vert < 1$
are complementary.
For the dichotomic field approach, Fig.~\ref{2015-qmusic-fig6} represents a configuration
of mutually complementary quantum tones for the note $a$ in the C major scale (a), and mutually complementary linear combinations as introduced in
Section~\ref{2021-qart-s2} (b).

Complementarity can be extended to more advanced configurations of contexts.
These quantum configurations and their associated quantum probability distributions,
if interpreted classically,
either exhibit violations of classical probability theory,
classical predictions, or nonisomorphic embeddability of observables into classical propositional
structures~\cite{Dzhafarov-2017,Abramsky2018,Grangier_2002,Auffeves-Grangier-2018,Auffves2020,Grangier-2020,cabello2021contextuality,specker-60,kochen1,svozil-2021-context}.

\section{Bose and Fermi model of tones}

An alternative quantization  to the one discussed earlier
is in analogy to some fermionic or bosonic---such as the electromagnetic---field.
Just as the latter one in quantum optics~\cite{glauber:70,glauber-collected-cat}
and quantum field theory \cite{Weinberg-search}
is quantized by interpreting every single mode
(determined, for the electromagnetic field  for instance by
a particular frequency and polarization)
as a sort of ``container''---that is,
by allowing the occupancy of that mode to be either empty or any positive integer (and a coherent superposition thereof)---we
obtain a vast realm of new musical expressions which cannot be understood in classical terms.

Whereas in a ``bosonic field model'' occupancy of field modes is easy to be correlated with the classical volume of the corresponding tone, in what follows we shall restrict ourselves to a sort of ``fermionic field model'' of music
which is characterized by a binary, dichotomic situation, in which every tone has either null
or one occupancy, represented by $\vert 0 \rangle= (0,1)$ or $\vert 1 \rangle = (1,0) $, respectively.
Thus every state of such a tone can thus be formally represented by entities
of a two-dimensional Hilbert space, $\mathbb{C}^2$, with the Cartesian standard basis
${\mathfrak B} =
\{
\vert 0 \rangle, \vert 1 \rangle
\}$.

Any note $\vert \Psi_i \rangle$ of the octave consisting of $\vert \Psi_c \rangle$,
$\vert \Psi_d \rangle$,~$\ldots$,
$\vert \Psi_b \rangle$,
in the C major scale
can be represented by the coherent superposition of its null and one occupancies; that is,
\begin{equation}
\vert \Psi_i \rangle =
\alpha_i \vert 0_i \rangle
+
\beta_i \vert 1_i \rangle
,
\end{equation}
with
$
\vert \alpha_i \vert^2 + \vert \beta_i \vert^2 =1
$, $\alpha_i. \beta_i \in \mathbb{C}$.

\begin{figure}
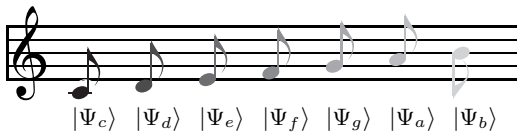

\begin{center}
\begin{music}
\startextract
\NOtes\zsong{$$}{\color{Black!0}\wh{d}}\en
\hspace{-10 mm}
\NOtes\zsong{$\vert \Psi_c \rangle$}{\color{Black!100}\ca{c}}\en
\NOtes\zsong{$\vert \Psi_d \rangle$}{\color{Black!85}\ca{d}}\en
\NOtes\zsong{$\vert \Psi_e \rangle$}{\color{Black!70}\ca{e}}\en
\NOtes\zsong{$\vert \Psi_f \rangle$}{\color{Black!55}\ca{f}}\en
\NOtes\zsong{$\vert \Psi_g \rangle$}{\color{Black!40}\ca{g}}\en
\NOtes\zsong{$\vert \Psi_a \rangle$}{\color{Black!30}\ca{h}}\en
\NOtes\zsong{$\vert \Psi_b \rangle$}{\color{Black!20}\ca{i}}\en
\zendextract
\end{music}
\end{center}
\caption{Temporal succession of tones $\vert \Psi_c \rangle$,
$\vert \Psi_d \rangle$,~$\ldots$,
$\vert \Psi_b \rangle$ in an octave in the C major scale with dicreasing mean occupancy.}
\label{2015-qmusic-fig2}
\end{figure}

Every tone is characterized by the two coefficients
$\alpha$ and $\beta$, which in turn can be represented (like all quantized two-dimensional systems)
by a Bloch sphere, with two angular parameters.
If we restrict our attention (somewhat superficially) to real Hilbert space $\mathbb{R}^2$,
then the unit circle, and thus a single angle $\varphi$,
suffices for a characterization of the coefficients $\alpha$ and $\beta$.
Furthermore, we may very compactly notate the mean occupancy of the notes by gray levels. Now, in this ``fermionic setting'', with the mean occupation number of any tone between 0 and 1 the gray level does not indicate the volume of the corresponding tone but the mere chance of it being present or not, see also Section~\ref{2021-qart-s2}.
Fig.~\ref{2015-qmusic-fig2} depicts a sequence of tones in an octave in the C major scale with decreasing occupancy,
indicated as gray levels.

In this case, any nonmonotonous unitary quantum musical evolution would have to involve the interaction of different tones;
that is, in a piano setting, across several keys of the keyboard.

\section{Quantum visual arts}

Just as for the performing arts such as music one could contemplate the quantum options and varieties for the visual arts.
Suffice it to say that the notion of ``color'' experience can be extended to the full quantum optical varieties
that result from the electromagnetic field quantization, as already mentioned earlier.
Incidentally, Schr\"odinger published a series of papers on classical color perception~\cite{Schrodinger1924,Schrodinger2017}
until around 1925.
Yet to our best knowledge he never considered the particular quantum aspects of human color and light perception.

Human rod cells respond to individual photons~\cite{Hecht1942,Westheimer2016}.
Moreover, recent reports suggest that humans might be capable of ``being aware'' of the detection of a single-photon incident on the cornea with a probability
significantly above chance~\cite{Tinsley2016}.
It thus may be suspected that this area of perception presents the most promising pathway into truly quantum perception. Speculations how this issue may be transferred to the perception of sound are compelling.

Let us state up front that quantum visual art, and, in particular, quantum parallelism, is not about additive color mixing,
but it is about the simultaneous existence of different, classically mutually exclusive ``colors'',
or visual impressions in general.
Quantum visual arts use the same  central assets or capacities (i)--(iv) mentioned earlier
in Section~\ref{2021-qart-s1}.
It can be developed very much in parallel to quantum music but
requires the creation of an entirely new nomenclature.
The perception of quantum visual art is subject to the same assumptions about the cognitive capacities to comprehend these
artifacts fully quantum mechanically or classically.
This will be shortly discussed in the following section.

\section{Can quantum art render cognitions and perceptions beyond classical art?}

Suppose for a moment that humans are capable to sense, receive and perceive
quantum signals not only classically but in a fully quantum mechanical way.
Thereby, they would, for instance, be capable of simultaneously ``holding'' different classically distinct tones at once---not
just by interference but by parallel co-existence.
This would result in a transgression of classical art forms,
and in entirely novel forms of art.

The existence of such possibilities depends on the neurophysiology of the human, or, more generally, sentient, perception
apparatus. Presently the question as to whether or not this is feasible is open; the answer to it is unknown.

In the case that merely classical perceptions are feasible, we would end up with a sort of Church-Turing thesis for music.
In particular, quantum music would not be able to ``go beyond'' classical music for a single observer,
as only classical renditions could be perceived.
Of course, as we mentioned earlier, quantum music might ``sound differently for different observers''.
To this end, we might conceptualize a kind of universal musical instrument that
is capable of rendering all possible classical notes.
Pianos and organs might be ``for all practical purposes good'' approximations to such a universal device.

Quantum music and quantum arts, just like quantum computing~\cite{deutsch}, or computations
starting and ending in real numbers but using imaginary numbers
as intermediaries~\cite{musil-toerless},  might be a sort of
bridge crossing elegantly a gap between two classical domains of perception.
And yet they could be so much more if only the quantum could be ``heard'' or ``sensed''.

\section{Summary}

We have contemplated the various extensions of music, and arts in general, to the quantum domain.
Thereby we have located particular capacities which are genuine properties.
These involve
parallelization through coherent superposition (aka simultaneous linear combination),
entanglement,
complementarity and
contextuality.
We have reviewed the nomenclature introduced previously~\cite{2015-qmusic}
and considered particular instances of quantum music.
Then we have briefly discussed quantum visual arts.

The perception of quantum arts depends on the capacity of the audience to either perceive
quantum physical states as such, or reduce them to classical signals.
In the first case, this might give rise to entirely novel artistic experiences.
We believe that these are important issues that deserve further attention, also
for sentient perception in general
and human neurophysiology, in particular.

\begin{acknowledgments}
This research was funded in whole, or in part, by the Austrian Science Fund (FWF), Project No. I 4579-N. For the purpose of open access, the author has applied a CC BY public copyright licence to any Author Accepted Manuscript version arising from this submission.
\end{acknowledgments}


%

\end{document}